\documentstyle[12pt]{article} \begin{document} \title{BE condensation
and independent emission: statistical physics
interpretation\thanks{Dedicated to Jan Pisut on the occasion of his 60th
birthday.}} \author{A.Bialas and K. Zalewski\thanks{Also at the Institute of
Nuclear Physics, Cracow.}
\\ M.Smoluchowski Institute
of Physics \\Jagellonian University, Cracow\thanks{Address: Reymonta 4,
30-059 Krakow, Poland; e-mail:bialas@thp1.if.uj.edu.pl;
zalewski@chall.ifj.edu.pl}}
\maketitle
\begin{abstract} Recent results on effects of Bose-Einstein
symmetrization in a system of independently produced particles are
interpreted in terms of statistical physics. For a large class of
distributions, the effective sizes of the system in momentum and in
configuration space are shown to shrink when quantum interference is
taken into account. \end{abstract}

\vspace{0.3cm}

{\bf 1.} In recent papers \cite{kz1,kz2} we worked out the implications of the
assumption that the identical pions generated in multiple production processes
have only the correlations due to Bose-Einstein statistics. This was taken to
mean that observable distributions can be evaluated in two steps. First, for
the unphysical case of distinguishable pions independence is assumed, i.e. the
poissonian multiplicity distirbution
\begin{equation}
P^{(0)}(N) =e^{-\nu} \frac{\nu^N}{N!}    \label{1}
\end{equation}
and the density matrix for each multiplicity in the form of  a product of
single particle
density matrices. In momentum representation, for $N$ particles we have
\begin{equation}
\rho^{(0)}_N(q,q')= \prod_{i=1}^N \rho^{(0)}(q_i,q_i'),   \label{2}
\end{equation}
where $q_i$ denotes  the momentum vector of  particle
$i$ and $q$ is the set of all $N$ momenta\footnote{The special case,
when $\rho^{(0)}$ is a gaussian,
has been studied already by a number of authors
 \cite{kz2}-\cite{zh}}. At this stage the momentum
distribution is given by the diagonal elements of the density matrix
\begin{equation}
\Omega^{(0)}_N(q) = \rho^{(0)}_N(q,q)        \label{3}
\end{equation}
and is normalized to unity
\begin{equation}
\int dq \Omega^{(0)}_N(q)=1.              \label{4}
\end{equation}
In the second step the distribution is symmetrized, i.e. the resulting momentum
distribution is obtained according to the prescription
\begin{equation}
\Omega_N(q) = \frac1{N!} \sum_{P,P'} \rho^{(0)} (q_P,q_P')
=\sum_P Re \rho^{(0)}(q,q_P),  \label{5}
\end{equation}
where the first sum runs over all permutations $P,P'$ of the momenta. As
seen from (\ref{5}), the momentum distribution $\Omega_N(q)$ is not normalized
to one any more. Consequently, the multiplicity distribution changes from
(\ref{1}) into
\begin{equation}
P(N)= C \frac{\nu^N}{N!} \int \Omega_N(q) dq          \label{6}
\end{equation}
where $C$ is a normalization constant.

We found for the generating function of the multiplicity distributinon
\cite{kz1} and for the generating functional for the multiparticle correlation
functions in momentum space \cite{kz2} explicit expressions in terms of the
eigenfunctions and eigenvalues of the single particle density matrix
$\rho^{(0)}$. In the present paper we describe a reinterpretation of
this model and show how the results can be simply obtained by using standard
statistical physics. This explains directly the physical meaning of the
results obtained in \cite{kz1,kz2}.

\vspace{0.3cm}

{\bf 2.}
The density operator corresponding to the density matrix $\rho^{(0)}(q_i,q_i')$
can be expressed in terms of its eigenvectors and eigenvalues
\begin{equation}
\hat{\rho}^{(0)} = \sum_n |n> \lambda_n <n| .      \label{7}
\end{equation}
Our first remark is that this operator corresponds to a single particle
canonical distribution, if we make the identification\footnote{We consider only
states with $\lambda_n > 0$. The states with $\lambda_n=0$ correspond to
infinite energy and do not play any role in our argument.}
\begin{equation}
\lambda_n= \frac1{Z} e^{-\beta \epsilon_n},      \label{8}
\end{equation}
where  $\beta = \frac1{kT}$, $\epsilon_n$ is the energy ascribed to the state
$n$, and $Z$ is the normalizing factor which ensures that the density operator
has the trace equal one
\begin{equation}
\sum_n \lambda_n =1.              \label{9}
\end{equation}
In terms of statistical physics, $Z$ is the single particle partition function:
\begin{equation}
Z=\sum _n e^{-\beta \epsilon_n}.          \label{10}
\end{equation}
The sum is over all states of the particle, thus a given energy $\epsilon_n$
can occur more than once. The hamiltonian with eigenstates $|n>$ and
eigenvalues $\epsilon_n$ is, of course,
\begin{equation}
H= \sum_n |n> \epsilon_n <n|.      \label{11}
\end{equation}
For instance, from results of  \cite{kz1} and \cite{kz2} we find that for the
Gaussian single particle density matrix in one dimension
\begin{equation}
\rho^{(0)}(q,q') = \frac1 {\sqrt{2\pi\Delta^2}}e^{-\frac{q_+^2}{2\Delta^2}
-\frac12 R^2 q_-^2},                                \label{12}
\end{equation}
where $\Delta^2$ and $R^2$ are positive constants constrained by the condition
$R\Delta \geq 1/2$ following from Heisenberg's uncertainty relation, the
Hamiltonian is that of a harmonic oscillator with
\begin{equation}
m\omega = \hbar\left (\frac{R}{\Delta}\right)^2                  \label{13}
\end{equation}
and
\begin{equation}
\hbar\omega = - kT \ln \frac{2R\Delta -1}{2R\Delta +1}.       \label{14}
\end{equation}
Inversely, for any Hamiltonian with a known set of eigenfunctions $\psi_n(q)$
and eigenvalues $\epsilon_n$ one can construct the corresponding density matrix
$\rho^{(0)}(q,q')$ using the formulae (\ref{7}) and (\ref{8}). This allows to
obtain a fairly  large class of explicitly solvable models.

Let us consider now a set of $N$ indistinguishable particles, where the
Hamiltonian is the sum over all the particles of single particle Hamiltonians
(\ref{11}). We assume that all these particles
have been produced independently\footnote{These assumptions correspond to the
mean field approximation in statistical physics. Thermodynamic equilibrium is
not assumed. For a discussion assuming thermodynamic equlibrium cf.
\cite{SIN}.} and that the probability of producing
$N$ particles is
$\bar{\nu}^N$. Let
us consider first the subset of particles in state $|n>$. Since the number of
particles in this subset is not fixed, we use the grand-canonical ensemble. The
probability of finding exactly $N$ particles in state $|n>$ is
\begin{equation}
P_n(N) = \frac1{{\cal Z}_n} \bar{\nu}^Ne^{-\beta N \epsilon_n}.   \label{15}
\end{equation}
The parameter $\bar{\nu}$, known as fugacity, is related to the chemical
potential $\mu$ by the formula
\begin{equation}
\bar{\nu}= e^{\beta \mu}   \label{16}
\end{equation}
and to $\nu$ by $\bar{\nu} = \nu/Z$. The normalizing factor
${\cal Z}_n$,  known as the
grand partition function, ensures that the sum of all the probabilities equals
one:
\begin{equation}
{\cal Z}_n= \sum_{N=0}^{\infty} \left(\bar{\nu} e^{-\beta \epsilon_n}\right)^N=
\frac1{1-\bar{\nu} e^{-\beta \epsilon_n}}.   \label{17}
\end{equation}
This formula makes sense only if the geometrical series is convergent, i.e. for
$\mu<\epsilon_n$.  The grand partition  function for the whole system is a
product of the grand partition functions for the independent subsystems, thus
it is
\begin{equation}
{\cal Z}=\prod_n \frac1{1-\bar{\nu} e^{-\beta \epsilon_n}}.   \label{18}
\end{equation}
The grand partition function contains much information about the system. In
particular it can replace the generating function for calculation of the
multiplicity distributions. For instance, the average number of particles in
state $n$ is
\begin{equation}
-kT\frac{\partial (ln{\cal Z}_n)}{\partial\mu} = \frac1{e^{\beta
(\epsilon_n-\mu)}-1}=\frac {\nu\lambda_n}{1-\nu\lambda_n}.  \label{19}
\end{equation}
The first form is the well-known Bose-Einstein distribution, the second is the
result obtained in \cite{kz1}. Thus the symmetrization according to formula
(\ref{5}) corresponds to the replacement of the Boltzman multiplicity
distribution by the Bose-Einstein one. For the probability of no particle in
state $n$ ($N_n=0$) we get from (\ref{15})
\begin{equation}
P_n(0) =\frac1{{\cal Z}_n} = 1- \bar{\nu}e^{-\beta\epsilon_n} =
\frac1{\bar{N}_n +1}.
\label{20}
\end{equation}
Two limiting cases are of interest here. When all the occupation numbers
$\bar{N}_i$ are very small $P(0)=\prod P_n(0) \approx e^{-\sum \bar{N}_n}
 = e^{-\bar{N}}$ is
very small for large $\bar{N}$. When almost all the particles
are in the ground state
(BE condensation), $P(0) \approx (1+\bar{N})^{-1}$ and this probability becomes
much larger.  Thus the probability of producing an event with no $\pi^0$'s
(centauro event \cite{fu}) is greatly enhanced by symmetrization, when a
significant fraction of particles is in the condensate. This happens when $\mu
\rightarrow \epsilon_0$ from below, or in terms of the eigenvalues of the
density matrix when $\nu\lambda_0 \rightarrow 1 $ from below.

\vspace{0.3cm}

{\bf 3.} The momentum distribution and correlation functions can be described
along the same lines. It was shown in \cite{kz2} that they can all be
expressed in terms of one function of two variables
$L(q,q')$ given by
\begin{equation}
L(q,q') = \sum_n \psi_n(q) \psi_n^*(q') \frac{\nu\lambda_n}{1-\nu\lambda_n}.
\label{21a}
\end{equation}
In particular, we have
\begin{equation}
\omega(q) = L(q,q)     \label{21b}
\end{equation}
for the inclusive single particle distribution, and
\begin{equation}
K_2(q_1,q_2) = |L(q_1,q_2)|^2     \label{21c}
\end{equation}
for the two-particle correlation function.

Using (\ref{8}) and (\ref{16}), Eq.(\ref{21a})  can rewritten as
\begin{equation}
L(q,q') = \sum_n \psi_n(q) \psi_n^*(q') \frac1{e^{\beta (\epsilon_n - \mu)} -1}
\label{21}
\end{equation}
Thus, it is just the element of the density matrix for the pure state $n$
averaged over the Bose-Einstein distribution. Replacing the Bose-Einstein
weights by the Boltzmann weights (\ref{8}) one recovers the unsymmetrized
single particle density matrix $\rho^{(0)}(q,q')$.  At this point let us
observe that the unsymmetrized density matrix $\rho^{(0)}(q,q')$ corresponds
better than the symmetrized one to the physical intuition since it
contains the parameters which are easier to intepret.
This rises an
interesting question, to what extent the apparent parameters of the system, as
determined from measured $\omega(q)$ and $K_2(q,q')$, are modified with respect
to the original "physical" parameters given by $\rho^{(0)}(q,q')$.

This question
was explicitly solved for the case of the Gaussian density
matrix (corresponding to the hamiltonian of a harmonic oscillator): the
resulting distributions are  narrower (both in momentum and in
coordinate space) than the original, unsymmetrized distributions \cite{kz2,zh}.
 It is not obvious, however,  how general this result is. Below we discuss this
problem for the  case of hamiltonians of the form
\begin{equation}
H = \frac{p^2}{2m} +V(x).   \label{22aa}
\end{equation}
with  $V(x)\rightarrow +\infty$ when $|x|\rightarrow \infty$, to ensure
the convergence of the trace of the density matrix.

To this end
let us first observe that the Bose-Einstein weights, when
compared with those of Boltzmann, enhance the contribution from states with
lower energy in the sum (\ref{21}). Thus,
 if the wave functions $\psi_n(x)$ and $\psi_n(q)$
broaden with increasing energies $\epsilon_n$ (decreasing $\lambda_n$),
the symmetrized distributions
are narrower than the unsymmetrized ones,
 i.e. we recover the qualitative result
 obtained for the Gaussian density matrix.
For such a situation to occur, it is  sufficient for example
 that the potential
$V(x)=\lambda |x|^{\alpha}$ where $\lambda$ and $\alpha$ are positive constants\footnote{The
authors thank V.Zakharov for calling their attention to the virial
theorem, which implies that.}.
Unfortunately, the necessary condition(s) are not so easy to determine.
It is also not immediately obvious,  what this
condition implies for the shape of the density matrix
$\rho^{(0)}(q,q')$.

 This argument also holds for the distribution in
configuration space, as determined from the correlation function
$K_2(q_1,q_2)$. In this case we  argue that the distribution in the
variable $q_1-q_2$ broadens when the symmetrization effects are introduced. To
this end let us observe that if all eigenvalues $\lambda_n$ were the same, the
sums in (\ref{21a}) and (\ref{21}) would be proportional to $\delta(q-q')$.
Thus $L(q_1,q_2)$ would be infinitely narrow in $q_1-q_2$, corresponding to an
infinite volume for particle production. Introducing the Boltzmann weigths
(\ref{8}) cuts the contribution from large energy levels and enhances the lower
ones. The result is broadening of $L$. Since, as we argued above, the
Bose-Einstein weigths work even stronger in the same direction, they will
broaden the distribution even more.

We thus conclude that the narrowing of the momentum and configuration space
distributions as a consequence of the symmetrization of the multiparticle wave
functions (first observed for the Gaussian density matrix) is actually a much
more general phenomenon, valid for a rather broad class of distributions. More
work is needed, however, to determine precisely the physical conditions which
determine the behaviour of the effective widths of the distributions after
symmetrization.

\vspace{0.3cm}

{\bf 4.} For fermions the difference is that $P_n(N) = 0$ for $N>1$. Thus the
formulae
(\ref{17}) and (\ref{18}) are replaced by
\begin{equation}
{\cal Z}_n^{(F)}= 1+ \bar{\nu} e^{-\beta\epsilon_n}; \;\;\;\;
{\cal Z}^{(F)} = \prod_n (1+ \bar{\nu} e^{-\beta\epsilon_n})   \label{22}
\end{equation}
The formula for the average population of state $n$ becomes
\begin{equation}
-kT\frac{\partial ln{\cal Z}_n}{\partial\mu} = \frac1{e^{\beta
(\epsilon_n-\mu)}+1}=\frac {\nu\lambda_n}{1+\nu\lambda_n}  \label{23}
\end{equation}
and, of course, there is no condensation. After symmetrization the elements of
 the density matrix $\rho^{(0)}$ are replaced by
\begin{equation}
L(q,q') = \sum_n \psi_n(q) \psi_n^*(q') \frac1{e^{\beta (\epsilon_n - \mu)} +1}
\label{24}
\end{equation}
One sees from (\ref{24}) that the Fermi-Dirac weights reduce the
contribution from the low-energy states as compared to the Boltzmann
weights which would be there without (anti) symmetrization of the wave
function. Thus the argument
 given in the previous section implies  that
 (anti)symmetrization of the wave function leads to broadening of the
fermion distribution both in momentum and configuration space.

\vspace{0.3cm}

{\bf 5.} In conclusion, we have shown that the assumption that the identical
bosons (fermions) generated in multiple production processes have only the
correlations due to Bose-Einstein (Fermi-Dirac) statistics can be reformulated
in the language of statistical physics. It corresponds to the standard
Bose-Einstein (Fermi-Dirac) distribution of particles in the mean field
approximation. Using this formulation we have shown for a large class of
distributions that symmetrization of the multiparticle wave function implies
narrowing of the particle spectra both in momentum and in configuration space
(broadening is expected in case of fermions). This generalizes the result found
earlier for the gaussian distributions \cite{kz2}-\cite{zh}.

\vspace{0.3cm}
{\bf Acknowledgements}
\vspace{0.3cm}

We would like to thank Piotr Bialas, Jan Pisut and  Andrzej
Staruszkiewicz for very useful discussions,
and to Ulrich Heinz for an interesting correspondence.
 This work was supported  by the KBN Grant No 2 P03B 086 14.

\end{document}